\documentclass[11pt,letterpaper,oneside,pdftex]{article}
\pdfoutput=1
\usepackage[utf8]{inputenc}
\usepackage[T1]{fontenc}
\usepackage{amsmath}
\usepackage{amsfonts}
\usepackage{amssymb}
\usepackage{graphicx}
\usepackage[width=16.20cm, height=22.50cm]{geometry}
\usepackage{fouriernc}

\usepackage[margin=1cm,font=small]{caption}
\usepackage[numbers,square,comma,sort&compress]{natbib}
\usepackage[pdftex,hyperref,svgnames]{xcolor}
\usepackage[pdftex,bookmarksnumbered=true,breaklinks=true]{hyperref}
\makeatletter
\AtBeginDocument{
	\hypersetup{
		pdftitle = {\@title},
		pdfauthor = {\@author}
	}
}
\makeatother
\makeatletter

\@addtoreset{equation}{section}
\makeatother

\newcommand{\vv}[1]{\mathbf{#1}}
\title{\texorpdfstring{\begin{flushright}
			{\small LA-UR-21-30714}
		\end{flushright}\vspace{2em}}{}%
Predicting electrical conductivity in Cu/Nb composites:
\texorpdfstring{\\}{}a combined model-experiment study}
\author{Daniel N. Blaschke, Cody Miller, Ryan Mier, Carl Osborn, Sean M. Thomas, \texorpdfstring{\\}{}Eric L. Tegtmeier, William P. Winter, John S. Carpenter, and Abigail Hunter}
\date{June 13, 2022}

\graphicspath{{./}{./figures/}}
\begin{document}
\maketitle

\thispagestyle{empty}
\begin{center}
	\vspace{-0.3cm}
	Los Alamos National Laboratory, Los Alamos, NM, 87545, USA
	\\[0.5cm]
	{\ttfamily{E-mail: dblaschke@lanl.gov, cmiller@lanl.gov, mier@lanl.gov, cmo@lanl.gov, smthomas@lanl.gov, etegtmeier@lanl.gov, wwinter@lanl.gov, carpenter@lanl.gov, ahunter@lanl.gov}}
\end{center}

% \vspace{1.5em}

\begin{abstract}
The generation of high magnetic fields requires materials with high electric conductivity and good strength properties.
Cu/Nb composites are considered to be good candidates for this purpose.
In this work we aim to predict, from theory, the dependence of electric conductivity on the microstructure, most notably on the layer thickness and grain sizes.
We also conducted experiments to calibrate and validate our simulations.
Bimetal interfaces and grain boundaries are confirmed to have the largest impact on conductivity in this composite material.
In this approach, a distribution of the layer thickness is accounted for in order to better model the experimentally observed microstructure.
Because layer thicknesses below the mean free path of Cu significantly degrade the conductivity, an average layer thickness larger than expected may be needed to meet conductivity requirements in order to minimize these smaller layers in the distribution.
We also investigate the effect of variations in volume fraction of Nb and temperature on the material's conductivity. 
\end{abstract}

\vspace{1cm}
\newpage
%\tableofcontents
%\newpage

\section{Introduction}

Creating ultra-high magnetic fields (e.g., 80-100T) are attractive and necessary to study fundamental questions and behaviors in a wide range of materials systems such as semi- and super-conductors, quantum matter and thin films.
Systems that can generate these high magnetic fields utilize conductive winding wire that must have a unique combination of material properties that encompass both high strength, to withstand Lorentz forces, and high conductivity \cite{Foner:1989,Campbell:1995,Embury:1998}.
In order to push the limit in attaining the highest possible magnetic fields, further improvement of the material properties in these wires is required.
The prevailing challenge in this regard has been the necessary trade-off between high electrical conductivity and high strength.

High strength two-phase composites fabricated via severe plastic deformation (SPD) methods such as accumulative drawing and bonding (ADB) or accumulative roll bonding (ARB), have been utilized to fabricate this winding wire for some time.
Many material candidates have been considered for these high-magnet applications, including Cu/stainless steel \cite{Dupouy:1995, Pantsyrnyi:2001}, Cu/Nb \cite{Dupouy:1995,Foner:1989,Pantsyrnyi:2001}, Cu/Cr \cite{Dobatkin:2015}, Cu/W \cite{Dong:2020}, Cu/Ta \cite{Zeng:2016}, and Cu/Ag \cite{Campbell:1995,Sakai:1991,Zhao:2016}.
Consequently, many studies have investigated microstructural effects on strength and conductivity properties of these composites in an effort to enhance the material properties \cite{Tsuji:2003,Shikov:2001,Ding:2021,Carpenter:2014,Carpenter:2022,Gu:2017,Vidal:2007,Dubois:2012,Rozhnov:2019,Ghalehbandi:2019,Gu2:2019}.
However, this remains a challenge as the material microstructure can significantly impact material properties in unexpected ways.
Specifically, different microstructural features can have opposing effects on strength and conductivity, making it difficult to optimize both properties.
For example, grain boundaries and/or bimetal interfaces can act as barriers to dislocation motion \cite{Zhang:2019,Kacher:2014,JWang2:2011}, increasing strength, while also scattering electrons \cite{Reiss:1986} and decreasing conductivity.

Cu/Nb wires have traditionally been of interest for high-magnet applications, and are commonly created using the ADB approach already mentioned \cite{Vidal:2007,Thilly:2000,Shikov:2001,Rozhnov:2019}.
Experimental studies have shown that these wire composites can achieve conductivity of $\sim$70\% IACS (where the International Annealed Copper Standard is by definition $5.8\times10^7$\,S/m) and an ultimate tensile strength, $\sigma_\text{UTS}$, of 1.1-1.3 GPa at 293 K \cite{Shikov:2001}.
Of course magnets operate over a wide range of temperatures ranging from 77--450 Kelvin \cite{Spencer:2004,Battesti:2018}.
At 77K Cu/Nb wires have exhibited $\sigma_\text{UTS}$=1.95 GPa and a low resistivity of 0.6 $\mu \Omega$ cm ($\widehat{=}\, 287$\% IACS) \cite{Thilly:2000}.
%%% 1/0.6e-8 = 16.667e7 S/m = 287\% IACS
More recently, ARB processed Cu/Nb composites have be studied for high magnet applications \cite{Ding:2021,Carpenter:2014,Carpenter:2022}.
Several studies have investigated strength in these composites as a function of layer thicknesses, finding that at room temperature, moderate layer thicknesses ($\sim$60 nm) have an $\sigma_\text{UTS}$ = 1 GPa \cite{Beyerlein:2013} and at relatively small layer thicknesses ($\leq$ 30 nm), the $\sigma_\text{UTS}$ can surpass 1 GPa \cite{Nizolek:2016}.
Strengths of nearly 1.8 GPa have been achieved with Cu/Nb ARB composites during micropillar compression at room temperature \cite{Beyerlein:2013} further illustrating the promising strength properties of these materials.
Fewer studies have addressed conductivity in these composites.
However recent measurements studied composites with 61\% IACS and a $\sigma_\text{UTS}$ = 1.2 GPa \cite{Ding:2021} and nearly 50\% IACS with $\sigma_\text{UTS}$ = ~750 MPa \cite{Carpenter:2022}, with the differences in these measurements due to differences in sample processing and volume fraction.

Clearly experimental studies have shown that these Cu/Nb composites have promising properties for high-magnet applications.
However, it is difficult to optimize both strength and conductivity properties with experiments alone.
Thus, there have also been several modeling efforts aimed at addressing these material systems.
While, to the authors' knowledge, there has not yet been a proposed modeling approach that has addressed both material strength and conductivity simultaneously, there have been several studies addressing microstructural effects on strength and conductivity separately. Strength behavior in ADB and ARB Cu/Nb materials has been studied extensively with different modeling approaches \cite{Gu2:2019,Hansen:2013,Jia:2013,Avallone:2019,TChen:2020,Shishvan:2021}.
Relatively fewer modeling studies have focused on conductivity, although there have been some addressing conductivity in both ADB Cu/Nb \cite{Tian:2014,HerveLuanco:2016,Gu:2017,Raabe:1995} and ARB Cu/Nb \cite{Ding:2021} systems specifically, along with other studies aimed at two-phase materials more broadly \cite{Lux:1993,Heringhaus:2003}.
Several of these approaches \cite{Ding:2021, Tian:2014} are based on Matthiessen's rule \cite{Matthiessen:1864}, which accounts for the degradation in material resistivity due to various scattering mechanisms, including the presence of interfaces, grain boundaries, dislocations, phonons, etc.
With the development of each of these terms, this model allows one to capture the well-known size-dependent resistivity, or size effect, that results in a sharp increase in resistivity (decrease in conductivity) when a characteristic dimension (e.g., layer thickness or wire diameter) is comparable to, or less than, the mean free path of the material \cite{Sondheimer:2001}.
We build upon this framework in the current study.

In particular, we present a combined experimental-modeling approach to investigate conductivity in ARB Cu/Nb composites.
Following the work of Tian et al. \cite{Tian:2014} and Jin et al. \cite{Jin:2013}, we present a model based on Matthiessen's rule that is implemented within a phase field framework.
Similar to previous work, we account for the effect of bimetal interfaces, grain boundaries, and static dislocation networks.
We inform these terms with experimental information, also generated as part of this work, in order to make as close a comparison to experimental conductivity data as possible.
We have also incorporated some additional features to better account for variations in the microstructure, such as the capability to capture a distribution of layer thicknesses rather than relying only on an average layer thickness.
We also note that most previous conductivity modeling studies have focused on Cu/Nb wire materials, rather than nanolaminate structures, with the exception of Ding et al. \cite{Ding:2021}.
With this modeling approach, we achieve very reasonable comparison to the experimental results, and also investigate the effect of temperature and volume fraction on the composite's conductivity.

This paper is organized as follows.
Section \ref{sec:macromodel} presents the details of the modeling approach.
The experimental approach, including conductivity measurements using the 4-point probe method, is described in Section \ref{sec:exps}.
Comparison between the simulated and experimental results are presented and discussed in Section \ref{sec:results}.
The work is then summarized, and main conclusions are discussed in Section \ref{sec:con}.

\section{Modelling Approach}
\label{sec:macromodel}

\subsection{Connecting microstructure and conductivity}

Experiments have shown that scattering of electrons on various crystal defects, solute atoms, etc. decreases conductivity, and phenomenologically it has been found that the effective resistivity can be written as a sum (Matthiessen's rule \cite{Matthiessen:1864})
\begin{align}
	\varrho = \varrho_0(T) + \varrho_d + \varrho_{gb} + \varrho_{ss} + \varrho_v +\varrho_{if}
	\,,
\end{align}
where $\varrho_0(T)$ is the temperature dependent phonon contribution, and $\varrho_d$, $\varrho_{gb}$, $\varrho_{ss}$, $\varrho_v$ are the contributions from dislocations, grain boundaries, solute atoms, and vacancies \cite{Lu:2004,Murashkin:2016,Chen:2019}.
Since this work is focused on modeling conductivity in ARB'd Cu/Nb composites, we also include a term for interfaces (specifically bimetal interfaces in this case), $\varrho_{if}$.
In addition, based on the Cu/Nb material of interest, we neglect the terms that describe decreases in conductivity due to the presence of solute atoms and vacancies.

The additional resistivity due to interfaces, $\varrho_{if}$, can be estimated with the following equation \cite{Dingle:1950,Tian:2014}:
\begin{align}
	\varrho_{if} &= \varrho_{0}(T)\left[1+\frac{3}{16}(1-p)\lambda_0(T)\frac{P}{S}\right]
	\,, \label{eq:rhointerface}
\end{align}
where $\varrho_{0}$ is the bulk resistivity, $p$ is the probability of elastic scattering at the interface, $\lambda_0$ is the bulk mean free path of electrons, and $P/S$ is the ratio of perimeter to cross-sectional area of the phase (or wire).
Note that this form, though often used for ADB materials, is nonetheless valid for any arbitrary cross section, and for a thin film with thickness $d_0$ the ratio tends to $P/S\to 2/d_0$, see \cite{Fuchs:1938,Dingle:1950}.

Typical values are $p\approx0.5$, see Ref. \cite{Heringhaus:2003}, and the mean free path, $\lambda_0$, ranges from a few up to a few tens of nano-meters for most metals, see Ref. \cite{Tian:2014}.
In particular, at room temperature $\lambda_0^\text{Cu}=39.9$nm for Cu according to Ref. \cite{Gall:2016}, and $\lambda_0^\text{Nb}=2$nm for Nb according to Ref. \cite{Tian:2014}.
These values are used to generate simulated results reported later in Section \ref{sec:results}.

Note that Eq. \eqref{eq:rhointerface} applies to deformation processed metal–metal composites (DMMCs), where both matrix and reinforcement metal phases have thicknesses perpendicular to the direction of current flow that is at least of the order (or larger than) the electron mean free path.
The scattering probability $p$ depends on the interface morphology; a rough interface tends to scatter inelastically and yields a low $p$ \cite{Fuchs:1938}.

Since, we are presently interested in body-centered cubic (bcc) layers (Nb) between layers of face-centered cubic (fcc) Cu, the ratio $P/S$ can be defined following Tian et al. \cite{Tian:2014} regarding volume fraction dependence, but using the thin film approximation of Fuchs \cite{Fuchs:1938,Dingle:1950} appropriate for our present flat ARB layers, i.e.:
\begin{align}
\frac{P}{S}\Big|_\text{Nb}\approx \frac{2}{d_0^\text{Nb}}
\,,
\end{align}
where $d_0$ is the average layer thickness.
Likewise,
\begin{align}
\frac{P}{S}\Big|_\text{Cu}\approx \frac{4V_f}{(1-V_f)d_0^\text{Cu}}
\,,
\end{align}
where $V_f$ is the volume fraction of Nb.
Additionally,
\begin{align}
\varrho_{0}(T) & \approx \frac{m_e}{n_e q_e^2\tau(T)}\,,
&
\lambda_0(T) & \approx \nu_F \,\tau(T)
\,,
\end{align}
with $m_e$, $q_e$ the mass and charge of an electron, and $n_e$ the concentration of electrons that are insensitive to temperature $T$ for an ideal free electron gas.
%% electron density n_e(Cu)\approx 8.49e28 electrons per m^3 %% (8.96e6/63.546)*6.022e23=8.4910e28 for 1 free electron per Cu-atom
%% electron density n_e(Nb)\approx 5.55e28 electrons per m^3 %% (8.570e6/92.90637)*6.022e23=5.5549e+28 for 1 free electron per Nb-atom (check this!)
$\tau(T)$ is the (possibly temperature dependent) relaxation time between scattering events and $\nu_F=\frac{\hbar}{m_e}(3\pi^2n_e)^{1/3}$ is the Fermi velocity. 

Grain boundary scattering becomes important when the grain sizes are small in the sense that their size is comparable to the electron bulk mean free path.
Following Ref. \cite{Tian:2014}, we estimate the additional resistivity due to grain boundaries as
\begin{align}
\rho_\text{GB} &= \rho_0(T)\left[\left(1-\frac32\alpha+3\alpha^2-3\alpha^3\ln\left(1+1/\alpha\right)\right)^{-1} - 1\right]\,,
\end{align}
where
\begin{align}
\alpha &= \frac{\lambda_0(T)}{d}\frac{R}{1-R}
\,,
\end{align}
$d$ is the average grain size, and model parameter $R$ is interpreted as a reflection coefficient of grain boundaries.
Recommended values in the literature are $R_\text{Nb}\sim 0.45$ for Nb \cite{Ding:2021,Tian:2014} and $R_\text{Cu}\sim 0.24$ for Cu \cite{Mayadas:1970}.
Note, however, that reflection coefficient $R$ depends on the grain size \cite{Feldman:2010} and shape geometry, and can in some cases take somewhat higher values, e.g. up to 0.83 for Cu \cite{Tian:2014}.

There have been claims in the literature that the scattering power of dislocations is so small that their effect on conductivity becomes important only for extremely high dislocation densities $\ge10^{17}$m$^{-2}$ \cite{Tian:2014,Karasek:1981,Karasek:1980}.
The resistivity due to dislocations is commonly parametrized as `scattering power' times dislocation density: $\varrho_d=R_d N_d$.
The recommended value for $R_d$ of Cu in the literature is 1.3$\times10^{-25}$ $\Omega\,$m$^3$ \cite{Brown:1977}, for example.
In Section \ref{sec:results} below we confirm that dislocation densities can be neglected for our present purposes.
In fact, we see that dislocation densities in our samples range from $10^{14}$--$10^{15}$\,m$^{-2}$ and in this range the change in conductivity is only a small fraction of a percent.

\paragraph{Introducing anisotropy}

In Ref. \cite{Dingle:1950} it was argued that the effective mean free path of electrons in a thin film takes the form
\begin{align}
	\lambda'(r_0) &= \lambda\left(1-e^{-r_0/\lambda}\right)
%	\approx r_0 + r_0^2/\lambda -\ldots
\end{align}
where $r_0=t/|\cos\theta|$, $t$ is the film thickness, and $\theta$ is the angle measured from the transverse direction.
Thus, for $\theta=0$ we have $r_0=t$ such that $\lambda'$ takes its smallest value.
Conversely, $\theta=\pi/2$ yields $\lambda'\to\lambda$.
We can use this generalization, which corresponds to introducing some anisotropy in a very simple fashion, to account for layers not being perfectly straight nor aligned with the direction of the electrical current (e.g., see the layer variation shown in the microstructure exhibited in Figure \ref{fig:layers}).
Simulations turned out not to be very sensitive to this effect unless the average layer thickness is close to the electron mean free path:
At 50nm layer thickness we observe a 3\% drop in conductivity if we allow $\theta$ to vary by up to 30 degrees using a sine function across the entire layer.
For layer thicknesses of 100nm, the change in conductivity drops to a fraction of a percent and we therefore neglect variations in $\theta$.

\subsection{Evolving the charge density within a phase field approach}

In this work, we evolve the charge density, and calculate the conductivity over time and space using a phase field formulation.
The specific phase field formulation used in this work was originally developed to evolve individual dislocations in order to study their interactions with each other and material microstructure in metallic materials \cite{Koslowski:2002,Beyerlein:2016}.
In this work we have implemented the means to calculate conductivity as a function of microstructure, however this formulation is not yet integrated with the dislocation evolution aspects of the phase field model.
The main advantage of integrating the conductivity calculation and the dislocation evolution is the future possibility of accounting for effects on dislocation motion such as electron drag, and also the corresponding effect on the overall material conductivity.
This integration is subject of current research and is the motivation for using a phase field framework.
Here we only discuss the conductivity calculation in detail, and interested readers may find more details about the dislocation evolution aspects of this phase field model in the following references \cite{Koslowski:2002,Beyerlein:2016,Albrecht:2020,Peng:2020,Smith:2020}.
In this case, the main advantage of using a phase field approach is that the gradients in conductivity (e.g. across interfaces) will be accounted for.

Following \cite{Jin:2013},
consider local charge density $\rho_e$.
Then the local current density $\vv{j}$ is given (via the microscopic version of Ohm's law) by
\begin{align}
	j_i &= \sigma_{ij} E_j(\vv{x})
	\label{eq:currentlocal}
\end{align}
where conductivity $\sigma$ may take a tensorial form to account for anisotropy.
The local electric field $\vv{E}$ is generated by the spatial distribution of charge density $\rho_e$ and the externally applied electric field, i.e.
\begin{align}
	\vv{E}(\vv{x}) &= \vv{E}^\text{ex} - \frac{i}{\varepsilon_0}\int \frac{d^3k}{(2\pi)^3} \frac{\tilde\rho_e(\vv{k})}{k^2}\vv{k}e^{i\vv{k}\vv{x}}
	\label{eq:Efieldlocal}
\end{align}
where $\tilde{\rho}_e$ is the Fourier transform of $\rho_e$.
As noted above, the law of charge conservation, $\dot\rho_e=-\nabla\vv{j}$, is related to the Ginzburg-Landau equations within a phase-field formulation.
In our practical application, we never need to write the energy explicitly, as we only need its change with time --- and that quantity is presently given by the divergence of the local electric current.
Upon choosing charge density $\rho_e$ as our new order parameter, the only unknown quantity is the local conductivity tensor $\sigma_{ij}$, for which we either need to have a model or experimental data (on a microscopic level).
Evolving the Ginzburg-Landau equations will give us a final charge density distribution, which via Eqs. \eqref{eq:Efieldlocal} and \eqref{eq:currentlocal} yields a current distribution $\vv{j\vv({x})}$.
The experimentally measured macroscopic conductivity is then determined by spatially averaging $\vv{j}$ and the macroscopic version of Ohm's law:
\begin{align}
	\langle j_i\rangle &= \sigma^\text{eff}_{ij} E_j^\text{ex}
	\,.
\end{align}
%%% Actually, perhaps we should not call this electron density but charge density: good initial conditions according to above paper would be rho=0.

Plugging Eqs. \eqref{eq:Efieldlocal} and \eqref{eq:currentlocal} into the charge conservation equation yields
\begin{align}
	j_i(\vv{x}) &= \sigma_{ij} E_j^\text{ex}  - \frac{i \sigma_{ij}}{\varepsilon_0}\int \frac{d^3k}{(2\pi)^3} \frac{{k}_j}{k^2}\tilde\rho_e(\vv{k}) e^{i\vv{k}\vv{x}}
\end{align}
%%% in the code we could store the Fourier transform of the second term above as jcurrent
and
\begin{align}
	\dot\rho_e&=-\partial_i\left(\sigma_{ij} E_j(\vv{x})\right)
	=-\partial_i\left(\sigma_{ij}  {E}_j^\text{ex} \right)
	-\frac{\sigma_{ij}}{\varepsilon_0}\int \frac{d^3k}{(2\pi)^3} \frac{k_i{k}_j}{k^2}\tilde\rho_e(\vv{k}) e^{i\vv{k}\vv{x}}
	+\frac{i}{\varepsilon_0}\left(\partial_i\sigma_{ij}\right)\int \frac{d^3k}{(2\pi)^3} \frac{{k}_j}{k^2}\tilde\rho_e(\vv{k}) e^{i\vv{k}\vv{x}}
	\,. \label{eq:LGeqConduct}
\end{align} 

In order to model local conductivity $\sigma_{ij}$, we start from the model presented in the previous section.
In the simplest case, where the conductivity is assumed isotropic and interactions with defects on a microscopic level are neglected, the present phase field approach will yield results which are very close to the macroscopic model of the previous section.
The reason is that the only term left to drive the conductivity to a different value is the gradient term in Eq. \eqref{eq:LGeqConduct}.

\section{Experimental Approach}
\label{sec:exps}

\begin{figure}[!htb]
\centering
\includegraphics[width=0.7\textwidth]{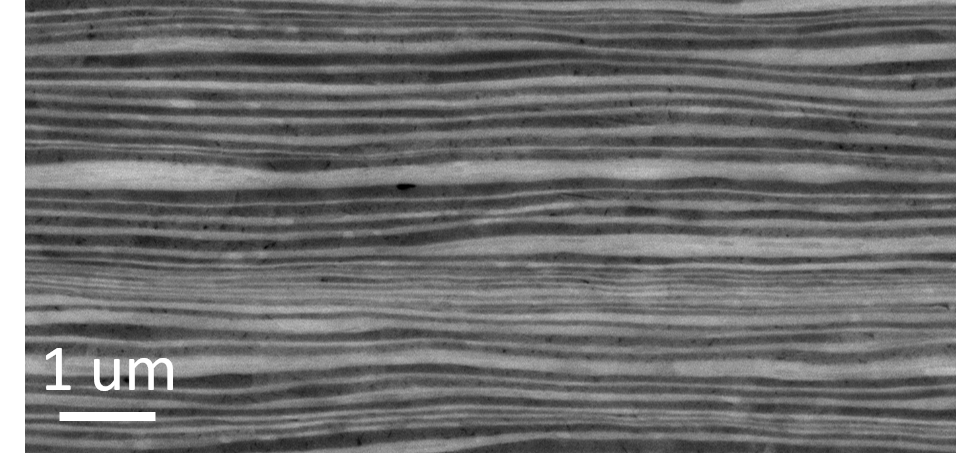}
\caption{Microscopy image showing the composite's multilayer structure.
Light layers are Nb and dark layers are Cu in this backscattered image.
Note that there is variation in layer thicknesses, but a layer intercept count method indicates that the nominal layer thickness of $h = 50$\,nm is within 5\% on average.}
\label{fig:layers}
\end{figure}

Rolling and roll bonding of material used in this study was performed on two rolling mills.
One mill is a two-high rolling mill (Waterbury-Farrel, Brampton, Ontario, Canada) with roll diameters of 406mm.
This instrument was used to process material via ARB to length scales greater than $d_0 = 150$nm.
Roll speeds of $\sim1.5$m/min were maintained with guides that reduced the amount of edge cracking.
Roll gaps were adjusted based on the overall height of the stacked materials.
No lubrication was utilized and no adjustments to the leading edge of the material were performed.
After roll bonding to $d_0 = 150$nm, a Stanat (Stanat, Long Island, NY) was used to further process the material at the nanoscale through rolling.
No further bonding was performed below $d_0 =150$nm but rolling with targeted reductions of 5\% were performed sequentially on unlubricated rolls that were held at room temperature.
This 2 high Stanat system had roll diameters of 150mm and rolling speeds of 4.5m/min were used.
Roll gaps were adjusted to target the 5\% reduction based on overall sheet thickness.

Starting materials used in this study were oxide free, high conductivity Cu that was purchased from Southern Copper and Supply and high purity Nb (99.97\% pure), which was purchased from ATI-Wah Chang.
The as-received Cu material was rolled to a thickness of 1 mm and a heat treatment under vacuum of 450$^\circ$C for one hour was performed.
The Nb material was rolled to a starting thickness of 2mm and a 950$^\circ$C one hour heat treatment under vacuum was performed.

Layer thicknesses of the samples were measured using a layer intercept count method on high energy backscattered electron images captured using a Thermo Fisher Helios G4 UXE Plasma FIB/SEM (Thermo Fisher Scientific, Massachusetts, USA).
Samples were mounted in order to allow investigation along the transverse direction of the rolled material.
Grinding and polishing procedures employed are covered in depth in \cite{Carpenter:2014}.
Note that the nominal layer thicknesses shared in Table \ref{tab:modexp} were confirmed via microscopy images (see Figure \ref{fig:layers}) with uncertainties of under 5\% after investigations with the layer intercept count method.

Conductivity was measured using the 4-point probe method \cite{Miccoli:2015}.
Dislocation density was quantified through a set of neutron diffraction experiments performed at the High Pressure Preferred Orientation (HIPPO) beamline at the Los Alamos Neutron Science Center at Los Alamos National Laboratory. 
Neutrons are able to provide a phase-based, bulk diffraction signal collected using several detector panel rings at multiple angles to provide near-3D diffraction pattern coverage that is integrated through the thickness of the Cu/Nb nanolamellar composites. 
A description of the HIPPO beamline can be found in \cite{Wenk:2003,Vogel:2004}. 
MAUD isotropic peak broadening model with an arbitrary texture assumption was used to extract phase specific dislocation densities \cite{Carpenter:2012,Takajo:2018}.

\section{Results}
\label{sec:results}
%%%%%%%%%%%%%%%%%%%%%%%%%%%

Here we present a direct comparison of experimental conductivity measurements and simulated conductivity predictions for Cu/Nb composites with different layer thicknesses and grain sizes.
In addition, using the model, we investigate the effect of variation in the volume fraction of Nb and temperature variations on the conductivity of the Cu/Nb composite.
The bulk conductivity at room temperature was measured in our samples as $\sigma_0^\text{Cu}=1/\rho^\text{Cu}_0 = 5.95\times10^7$ S/m for Cu
and $\sigma_0^\text{Nb}=1/\rho^\text{Nb}_0 = 6.334\times10^6$ S/m for Nb.
The naive mean value for the conductivity of the composite (without interface and other effects) is
$\sigma^\text{mean} = \sigma^\text{Nb}V_f+(1-V_f)\sigma^\text{Cu}$,
where $V_f$ denotes the volume fraction of Nb.

\subsection{Effect of microstructure on conductivity}

The presence of bimetal interfaces in the Cu/Nb composite is expected to have the largest scattering impact, and thus will have the greatest effect on the material's conductivity \cite{Tian:2014,Ding:2021}.
Table \ref{tab:modexp} presents both simulated and experimental results for conductivity of the 50/50 volume fraction Cu/Nb composite as a function of layer thickness.
In this case, the simulated results only take into account the effect of bimetal interfaces.

Table \ref{tab:modexp} also presents the results from the naive mean, which does not change with layer thickness and essentially presents an ideal value for the conductivity since no scattering effect from the microstructure is accounted for.
As expected, this value is much higher than both the simulated and experimental values.
Conversely, the experimental results show an effect of the change in layer thickness, with the conductivity generally increasing as the layer thickness increases, particularly from 50nm to 100nm.
The simulated results, in which the effect of the bimetal interfaces are accounted for, also follow this general trend. 

The comparison between the simulated results, $\sigma^\text{if}$, and the experimental results are reasonable, however the model consistently predicts higher conductivity values with respect to the experimental values in all cases.
The effect of bimetal interfaces does have a significant impact on the material conductivity, and is notable even at the large layer thickness of 500nm.
We note that our simulation results were stable with respect to the number of layers simulated, i.e. we got the same result for 4 layers each of Cu and Nb as for 8 layers each.

\begin{table*}[ht]
{\renewcommand{\arraystretch}{1.2}
\centering
\begin{tabular}{c|c|c|c|c|c}\hline
$V_f$  & $d^\text{Nb}_0$[nm] & $\sigma^\text{mean}$[$10^7$S/m] & $\sigma^\text{if}$[$10^7$S/m] & $\sigma^\text{exp}_\text{4p}$[$10^7$S/m] & \%err$_\text{4p}$ \\\hline
0.5 & 50 & 3.2917 & 2.6041 & 2.56 & 1.72 \\
0.5 & 100 & 3.2917 & 2.9033 & 2.67 & 8.74 \\
0.5 & 150 & 3.2917 & 3.0211 & 2.64 & 14.44 \\
0.5 & 200 & 3.2917 & 3.0840 & - & - \\
0.5 & 500 & 3.2917 & 3.2050 & - & -
\end{tabular}
\caption{Comparison of simulated and experimentally determined conductivity for 50/50 Cu/Nb material with varying average layer thickness.
Modeling results show the mean value (no effect due to microstructure) and the value that accounts for the effect of bimetal interfaces ($\sigma^\text{if}$).
% All values are given in SI units.
The layer thicknesses of the Cu phase are implied: $d_\text{Cu}=d_\text{Nb}\left(1-V_f\right)/V_f$.
In the last column we show the errors relative to the
% two experimental results using two different measurement techniques: the 4-point probe and the eddy current method.
experimental results.
}
\label{tab:modexp}
}
\end{table*}

Earlier work \cite{Carpenter:2012b,Carpenter:2013,Zeng:2017,You:2021} has shown that the majority of (especially thinner) layers consist of only one grain in the transverse direction, with grains significantly elongated in the rolling direction.
Typical grain lengths are of the order of 1$\mu$m.
We therefore do not take into account any grain boundary effects on the conductivity when the Cu/Nb layers are smaller than 60nm.
However, as layer thickness increases, multiple grains may be present across the transverse direction in each layer, and we would expect these interfaces would have the next largest scattering effect in comparison to the effect from the bimetal interfaces.
This is one likely reason for an increase in the \% error as the layer thickness increases shown in Table \ref{tab:modexp}.
We note that we neglect grain boundary segments that are perpendicular to the bimetal interface as they are relatively small segments, limited by the layer thickness, that are parallel to the current flow and sparse since grains are elongated in the rolling direction.

Thus, for thicker layers, we assume a small number of grains in the transverse direction and incorporate the effect of these boundaries into our conductivity calculations.
We account for grain boundaries in a weighted fashion, considering that a layer with $k$ grains in the transverse direction has $k-1$ grain boundary interfaces and 2 bimetal interfaces where the layer is bonded to the other material on either side.
Table \ref{tab:modelling2} shows the effect of grain boundaries on the conductivity in addition to bimetal interfaces for 50/50 Cu/Nb material with the largest layer thickness tested, 500nm.

\begin{table*}[ht]
{\renewcommand{\arraystretch}{1.25}
\centering
\begin{tabular}{c|c|c|c|c|c|c}\hline
$V_f$  & $d^\text{Nb}_0$\,[nm] & $d^\text{Nb}_{G}$\,[nm] & $d^\text{Cu}_{G}$\,[nm] & $\sigma^\text{mean}$\,[$10^7$S/m] & $\sigma^{\text{if+gb}}_{R_\text{Cu}=0.24}$[$10^7$S/m] & $\sigma^{\text{if+gb}}_{R_\text{Cu}=0.83}$[$10^7$S/m] \\\hline
0.5 & 500 & 250 & 250 & 3.2917 & 3.1632 & 2.4715\\
0.5 & 500 & 100 & 250 & 3.2917 & 3.1592 & 2.4675\\
0.5 & 500 & 250 & 100 & 3.2917 & 2.9440 & 1.3869\\
0.5 & 500 & 100 & 100 & 3.2917 & 2.9401 & 1.3829\\
\end{tabular}
\caption{Modeling results showing the effect of grain boundaries and bimetal interfaces (last two columns).
% All values are given in SI units.
The layer thicknesses of the Cu phase are implied: $d_\text{Cu}=d_\text{Nb}\left(1-V_f\right)/V_f$.
Furthermore, $d^\text{Nb}_{G}$ and $d^\text{Cu}_{G}$ denote the average grain sizes of Nb and Cu.
Thus, the number of grains per layer in the transverse direction is implied, e.g.  $d^\text{Nb}_0 / d^\text{Nb}_{G}$ for the Nb layers and $d^\text{Cu}_0 / d^\text{Cu}_{G}$ for Cu layers.
}
\label{tab:modelling2}
	}
\end{table*} 

Recall that for the 50/50 500nm layer cases without grain boundaries, the model predicted a conductivity of 3.2050e7 S/m.
As expected, accounting for the presence of grain boundaries does notably degrade this value, as seen in Table \ref{tab:modelling2}.
Of course the more grains present across a layer's thickness, the more significant the degradation in conductivity.
However, Table \ref{tab:modelling2} shows that the presence of grain boundaries in Cu, has a more significant impact on the conductivity of the composite than grain boundaries present in Nb.
This is because Cu has an order of magnitude higher bulk conductivity than Nb.
Since literature values for the grain boundary reflection coefficient $R_\text{Cu}$ of Cu range from 0.24 \cite{Mayadas:1970} to 0.83 \cite{Tian:2014}, with the exact value depending on shape and size of the grains \cite{Feldman:2010}, we show in Table \ref{tab:modelling2} results for both $R_\text{Cu}=0.24$ and $R_\text{Cu}=0.83$ to highlight the importance of knowing the grain size and shape distribution in order to improve the theoretical prediction of conductivity.

Figure \ref{fig:layers} shows that not all layers in a sample were equally thick, rather we always observed a spread in layer thickness.
Thus, the layer thicknesses reported in Table \ref{tab:modexp} are the \emph{average} layer thicknesses.
This is another likely reason that the calculated conductivity values in Table \ref{tab:modexp} are larger than the experimental values for a given layer thickness.
To understand the effect of having a variation in layer thicknesses within a sample, we computed from our model conductivities for layer thicknesses ranging from 25nm to 300nm in 25nm increments.
We then generated a Gauss distribution with $10^6$ samples and a standard deviation of 40nm around the mean layer thickness (cutting off any values below 25nm or above 300nm).
From these distributions, a mean conductivity value was computed.

\begin{figure}[!htb]
\centering
\includegraphics[width=0.75\textwidth]{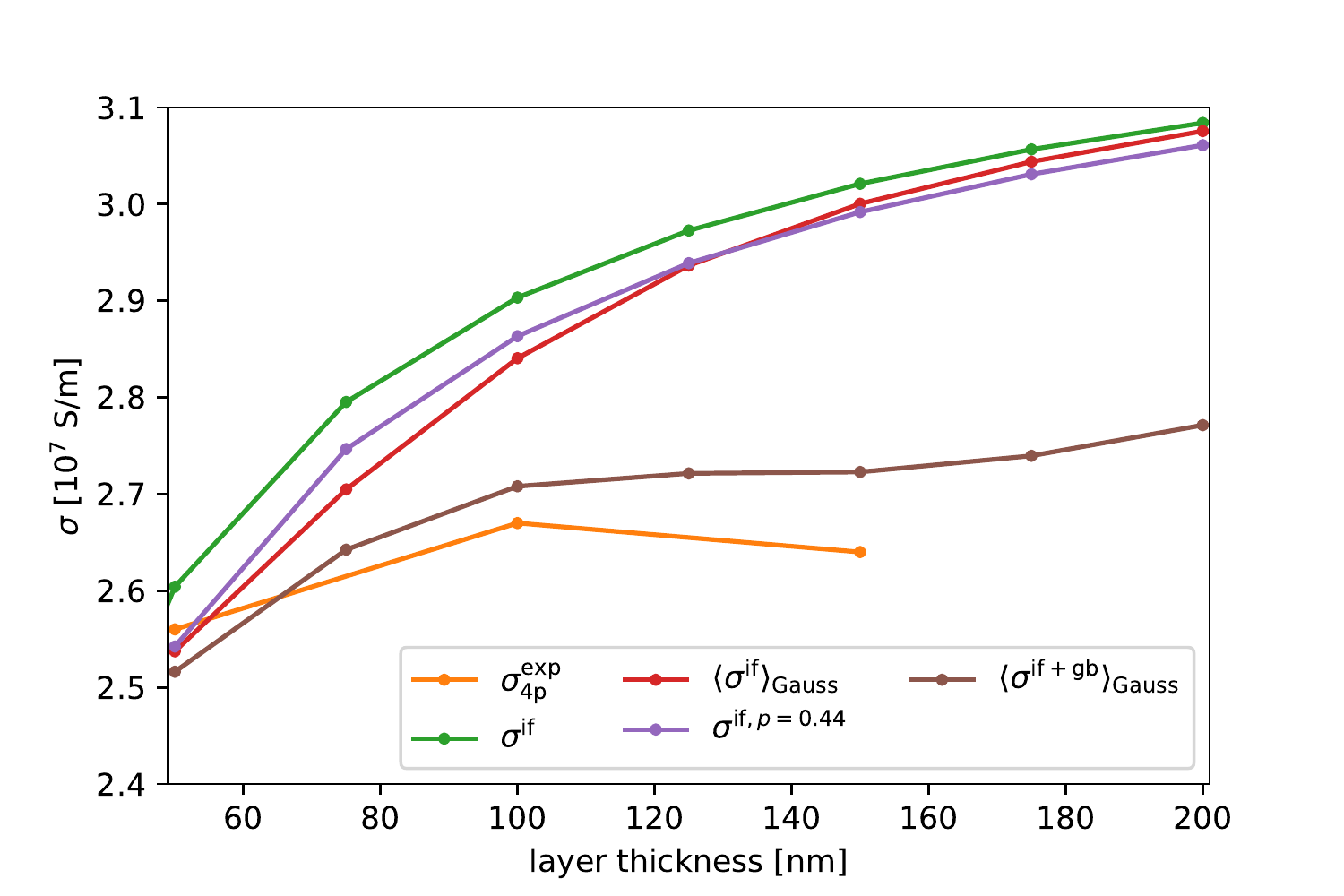}
\caption{We show electric conductivity as a function of layer thickness for a Cu/Nb composite.
The orange curve shows our experimental results using the 4-point probe method.
The green curve shows model results taking into account only the effect of interface scattering.
The red curve shows the effect of having a distribution of layer thicknesses and was computed using a Gauss distribution of the results shown in green with a Standard deviation of 40nm.
The purple curve shows that tuning the model parameter $p$,
which encodes the efficiency of scattering electrons at the bimetal interfaces, also reduces the electric conductivity, but in a more homogeneous fashion than the effect of having a spread in layer thickness.
The value for $p$ was reduced from 0.5 to $p=0.44$ so as to roughly match the 50 nm value of the red curve.
For the brown curve, we have assumed 25\% of the thicker layers ($\ge125$nm) have 2 grains across the layer thickness in addition to having a distribution of layer thicknesses with a Standard deviation of 40nm.
The grain boundary reflection coefficient was assumed at its highest value, 0.83, for demonstration purposes.}
\label{fig:expvssimul_spread}
\end{figure}

Results from these computations are shown in comparison to the experimental values and computed values that do not account for any layer thickness variation in Figure \ref{fig:expvssimul_spread}.
When variation of layer thickness is accounted for, the calculated conductivity values decrease, approaching the experimental values.
The effect is more significant at small layer thickness ($\sim$100nm and below).
This is due to the fact that as the average layer thickness decreases, the likelihood of layers smaller than the mean free path in Cu ($\sim$40nm) increases.
These layers will have a substantial impact on the overall conductivity of the Cu/Nb composite.

In that same figure, we also show that in tuning the model parameter $p$, which encodes the efficiency of scattering electrons at the bimetal interfaces, the electric conductivity is also reduced, but in a more homogeneous fashion than the effect of having a spread in layer thickness.
Differences in layer distributions and number of grains across the layer thickness are likely contributing factors to the slightly lower conductivity measured at 150nm average layer thickness vs the value measured at 100nm.
Thus, it is important to note that the model predictions could be improved if the maximum/minimum layer thickness and standard deviation could be extracted from the experimental data and used to inform the model.
However, at present extracting such information is quite difficult, but new data analysis tools may make it easier to determine such information in the future.

For demonstration purposes, we show an additional curve in Figure \ref{fig:expvssimul_spread}, where we account for multiple grains per layer thickness in a small subset of (thicker) layers.
In particular, in order to improve agreement with the experimental values, we assumed 25\% of the layers $\ge125$nm had 2 grains across the thickness and assumed the grain boundary reflection coefficient of Cu at its highest literature value of 0.83.
A higher reflection coefficient has a similar effect as assuming more (smaller) grains across the thickness.
The lowest value (0.24), did not have a large enough effect to bring down the 150nm conductivity value if we assume grains are at least 60nm thick.
Our assumptions regarding the percentage of layers with 2 grains across the thickness and the minimum layer thickness for considering such multigrain layers, though somewhat supported by experimental data \cite{Carpenter:2012b,Carpenter:2013,Liu:2014}, have been tuned to match the experimentally determined conductivity shown in orange within Fig. \ref{fig:expvssimul_spread}.
Obviously there are several parameters that can be tuned, unless we have exact measurements regarding grain size and number distribution.
In principle, the grain distribution within thicker layers can (and likely will) be different.
The main point of this discussion is to explain the somewhat unexpected drop in conductivity observed at 150nm.
A more accurate prediction, however, will required more detailed information about the microstructure, most notably layer thickness and grain size distribution within the bimetal.

Finally, our modeling approach can account for the effect of a (static) dislocation density within the Cu/Nb samples.
When including this feature, our simulations confirm claims in the literature that a dislocation density of $10^{15}$\,m$^{-2}$ or less has a negligible effect on conductivity.
In particular, a dislocation density of $10^{15}$\,m$^{-2}$ lowers the composite's conductivity by 0.52\% if all layers are 50nm thick and by 0.65\% for 300nm layers.
Increasing the dislocation density to $10^{16}$\,m$^{-2}$ lowers the conductivity by 5.94\% for 50nm layers and by 6.17\% for 300nm layers.
Once the dislocation density reaches $10^{17}$\,m$^{-2}$, conductivity is lowered by more than 30\%.
Since experimental measurements of our samples have revealed dislocation densities in the range $10^{14}$--$10^{15}$ m$^{-2}$, we neglect the dislocation density effect in our simulations presented here.
We note that these dislocation density measurements are comparable, but less than what is seen in Cu/Nb wire composites, which report dislocation densities of $10^{16}$--$10^{17}$ m$^{-2}$ \cite{Karasek:1980,Karasek:1981}.
Thus, in these wire composites the degradation of conductivity due to the dislocation network would be significant, particularly considering the maximum dislocation density value of $10^{17}$ m$^{-2}$, and this term would be required in the model to generate reasonable predictions.
This corresponds to high resistivity due to high dislocation densities observed in experiments in Cu/Nb wire composites \cite{Karasek:1980}.

\subsection{Temperature and volume fraction dependence of conductivity}

\begin{figure}[!htb]
\centering
\includegraphics[width=0.7\textwidth]{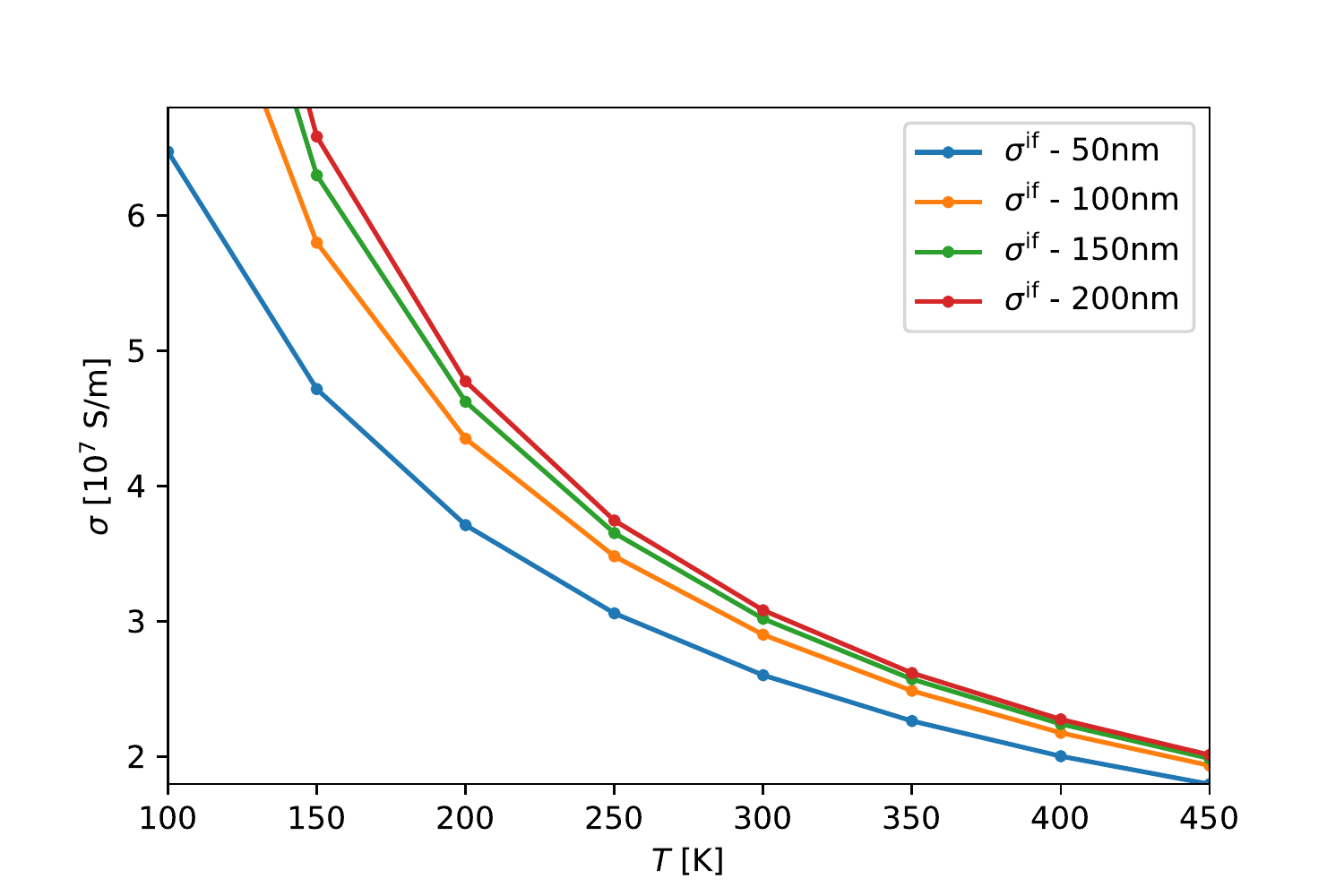}
\caption{We show electric conductivity as a function of temperature for a 50/50 Cu/Nb composite with different layer thicknesses.}
\label{fig:Tdep}
\end{figure}

\begin{figure}[!htb]
\centering
\includegraphics[width=1.\textwidth]{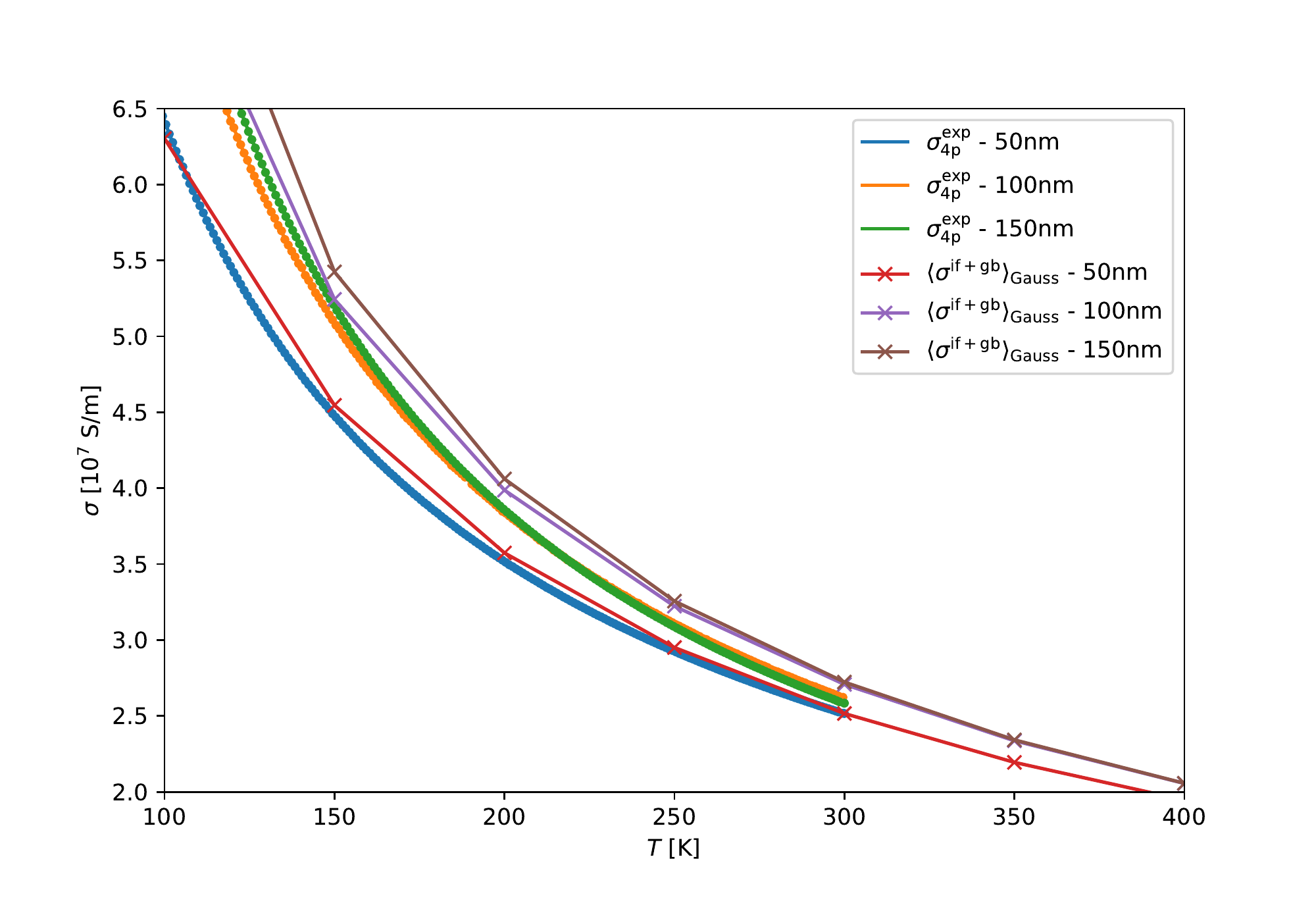}
\caption{We compare our simulation results for electric conductivity as a function of temperature for a 50/50 Cu/Nb composite with different layer thicknesses to our experiments using the 4-point probe method.
The processing temperature during ARB was 25$^\circ$C.
In the simulations shown here, we also accounted for a distribution in layer thickness as described in the previous section.
We assumed that 25\% of all layers with thicknesses of $\ge125$nm within that distribution contained two grains across the layer thickness.
}
\label{fig:Tdep+exp}
\end{figure}

Since our main motivation for the present study is to optimize Cu/Nb composite metals for magnet applications, we also investigate how the conductivity of the layered material depends on temperature.
In particular, we are interested in temperatures in the range 80--450 Kelvin \cite{Spencer:2004,Battesti:2018}.
Hence, we show in Figure \ref{fig:Tdep} the Cu/Nb composite's conductivity as a function of temperature.
In this figure, we have assumed the bulk resistivity changes linearly with temperature ($\Delta\rho_\text{Cu}=0.385$\% and $\Delta\rho_\text{Nb}=0.318$\% per Kelvin), which in the temperature range 150K--450K is a good approximation with deviations to the true bulk conductivity values for Cu being less than 3\%, see Ref. \cite{CRChandbook}.
At 100K the linearly extrapolated bulk conductivity in Cu is about 11\% too low, but with the additional resistivity from the bimetal interfaces and the Nb that error is reduced to about 3.5\%--6.5\% in the final prediction for the composite with layer thicknesses ranging from 50nm--200nm.
Once more, we accounted for only one grain across the layer thickness.
The non-linear behavior observed in this figure is due to the second term in the layer interface scattering model, Eq. \eqref{eq:rhointerface}, where the electron mean free path $\lambda_0(T)$ changes with temperature by the same rate as the bulk resistivity $\rho_0(T)$.

The CRC handbook, Ref. \cite{CRChandbook}, cites a bulk resistivity value of 0.215$\mu\Omega$\,cm for Cu at 80K.
Using this value within our simulation for layer thicknesses of 200nm, we calculate an effective conductivity for the Cu/Nb composite of $\sigma^\mathrm{if}=15.659\times10^7$S/m, which is comparable to the conductivity Ref. \cite{Thilly:2000} found for an ADB Cu/Nb wire at 77K (their measured resistivity value being $\rho=0.6\mu\Omega$\,cm, which corresponds to a conductivity of $\sigma=16.667\times10^7$S/m).
Of course this comparison is only qualitative, since the Cu/Nb wire of Ref. \cite{Thilly:2000} was processed using ADB with Nb fibers within a Cu matrix, whereas we presently discuss ARB Cu/Nb composites.
Furthermore, this particular simulation assumed fixed 200nm layer thicknesses for both materials, i.e. a 50\% volume fraction.
Thus, the comparison merely serves to show that our simulations predict reasonable values.

In subsequent Figure \ref{fig:Tdep+exp}, we compare our simulation results to values we measured.
We slightly overpredict conductivity for average layer thicknesses $\ge100$nm.
At 50nm average layer thickness, the agreement is the best, as expected.
In these simulations we also accounted for a distribution in layer thickness as described in the previous section.
We assumed that 25\% of all layers with thicknesses of $\ge125$nm within that distribution contained two grains across the layer thickness.
For the latter grain boundary model, we used the higher value of $R_\text{Cu}=0.83$ as recommended in Ref. \cite{Tian:2014}.

\begin{figure}[htb]
\centering
\includegraphics[width=0.7\textwidth]{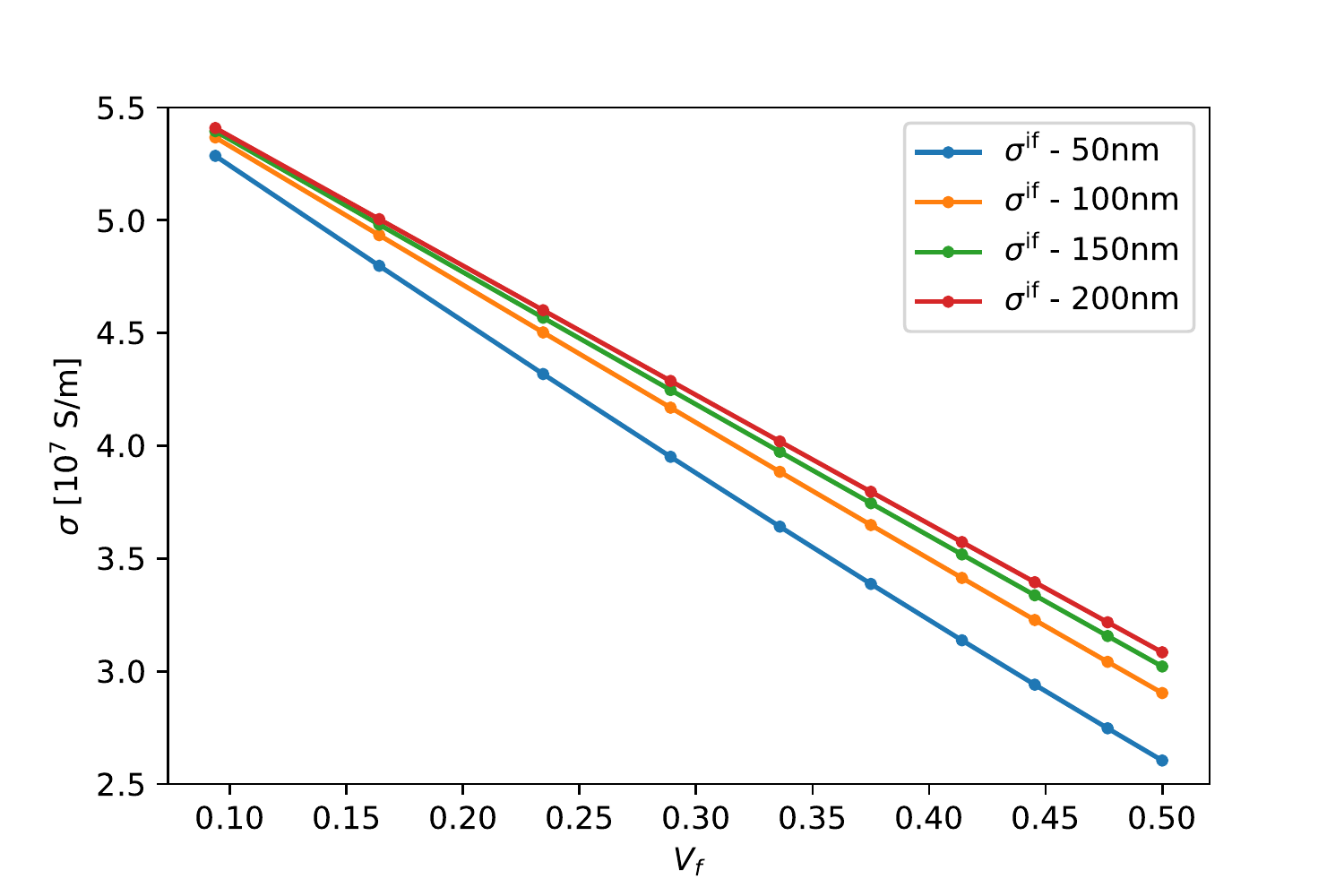}
\caption{We show electric conductivity as a function of volume fraction for a Cu/Nb composite with different layer thicknesses.
The Cu layers were kept at the fixed values shown in the legends, while the Nb layer thicknesses varied according to the x axis.
The Nb volume fraction thus computes as $V_f=d_\text{Nb} / \left(d_\text{Nb}+d_\text{Cu}\right)$.}
\label{fig:Vf}
\end{figure}

Finally, using this model, we can investigate the effect of varying the Cu/Nb volume fraction to see if perhaps there are volume fractions that provide better conductivity than 50/50.
For these simulations, we kept the Cu layers at fixed values and varied the thickness of the Nb layers.
Results are presented in Figure \ref{fig:Vf} for 4 different thicknesses of Cu layers, shown in the figure legend.
At volume fraction $V_f=0.5$, the Nb layer thicknesses are equal to the Cu layer thicknesses, but then decrease with decreasing volume fraction.
As expected, the volume fractions that are predominantly Cu produce the highest conductivity, however such composites would not have the necessary strength needed to be useful in magnet applications.
Past strength measurements of Cu/Nb composites have focused only on a few specific Nb volume fractions \cite{Han:1999,Ding:2021} where it was shown that the composite is stronger than its constituents \cite{Bevk:1978}.
Ding et al. \cite{Ding:2021} recently achieved a yield strength of 1.06 GPa and an ultimate tensile strength of 1.2 GPa together with a conductivity of $\sim61$\% IACS ($\widehat{=}\,3.538\times10^7$\,S/m) in a Cu/Nb composite with Nb volume fraction of 1/3.
A thorough experimental survey of strength as a function of volume fraction over a wide range of the latter would be useful in order to determine the permissible volume fraction of Nb necessary to meet strength needs in magnet applications and to find the ``sweet spot''.

Our simulated conductivity results as a function of volume fraction then show a nearly linear decrease in conductivity as the amount of Nb increases in the composite. 
50/50 volume fraction composites present the lowest conductivity.
The layer thickness of Cu has very little effect for Cu layers of 100nm and larger.
However, when the Cu layers are 50nm, there appears to be additional degradation in the conductivity starting at Nb volume fractions of about 0.2.

\section{Conclusion and Outlook}
\label{sec:con}
%%%%%%%%%%%%%%%%%%%%%%%%

In this paper, we studied the electrical conductivity of Cu/Nb composite metals using a combined modeling and experimental approach.
Several aspects of the ARB'd composite's microstructure were considered to understand both the features that have a dominate effect on conductivity, and also to develop a predictive model framework.
In particular, we investigated how not only the volume fraction of Nb, but also variation in layer thickness and the number of single crystal grains across the layer thickness, influences the conductivity.
Among the different microstructural effects, bimetal interfaces and grain boundaries are confirmed to have the largest impact on conductivity.
We also studied the impact temperature variation had on our conductivity results within the range 100K--450K.
Overall, our simulations compare well to the experimental results presented here with regard to average layer thickness and temperature dependence.
Thus, this model is a viable approach for exploring conductivity in nanolaminate composites over a range of temperatures and sample parameters (i.e., layer thicknesses and volume fractions), which can be valuable for determining promising compositions that warrant further experimental testing.
Our main advances compared to previous work of others, such as Ref. \cite{Ding:2021}, are:
\begin{itemize}
\item We account for a distribution of layer thicknesses in our simulations, which better matches the real material microstructure, since experiments show that layer thickness can vary substantially within a single sample.
Layer thicknesses below the mean free path of Cu significantly degrade conductivity.
Thus, to optimize conductivity, one may actually need a larger average layer thickness than expected to minimize the presence of these small layers in the distribution.
\item We consider temperature dependence and present a complete experiment / simulation study where we show that, upon accounting for a distribution in layer thickness and the presence of grain boundaries in larger layers, that our simulations match the experimental results reasonably well over a wide range of temperatures.
\item We studied the influence of varying the volume fraction of Nb within the Cu/Nb composite on the overall conductivity within our simulations.
Additional measurements for volume fractions other than 0.5 are subject of future work.
\end{itemize}

While the presence of dislocations does not have a significant effect on the conductivity, as we have confirmed here, the presence of an electrical field could impact the motion and interaction of dislocations via the `electron wind force' \cite{Fiks:1981,Antolovich:2004,Hariharan:2017}, which could impact a material's strength behavior.
Incorporating this is left for future work, however the phase field approach utilized in this work, which can resolve individual dislocations, opens the door for a model that can capture coupled dislocation-electrical field dynamics now that conductivity has been successfully incorporated.
This would result in a computational model that could capture conductivity and material strength simultaneously for the first time.

\subsection*{Acknowledgments}
%%%%%%%%%%%%%%%%%%%%%%%%%%%%%%%%%%%%%%%%%%

This work was supported by the U.S. Department of Energy through Los Alamos National Laboratory.
Los Alamos National Laboratory is operated by Triad National Security, LLC, for the National Nuclear Security Administration of the U.S. Department of Energy under contract 89233218CNA000001.
This work was funded through Los Alamos National Laboratory Directed Research and Development (LDRD) project  ER20200375.
S.M. Thomas acknowledges support from the U.S. Department of Energy, Office of Basic Energy Sciences, Division of Materials Science and Engineering project ``Quantum Fluctuations in Narrow-Band Systems''.
We also thank F. Ronning for related discussions.

%%%%%%%%%%%%%%%%%%%%%%%%%%%%%%%%%%%%%%%%%%%
\bibliographystyle{utphys-custom}
\bibliography{microstructure}

\providecommand{\accepted}[1]{accepted for publication in \textit{#1}}
\providecommand{\href}[2]{#2}\begingroup\begin{thebibliography}{10}
\small\itemsep=3pt
\tolerance 1414
\hbadness 1414
\emergencystretch 1.5em
\hfuzz 0.3pt
\widowpenalty=10000
\vfuzz \hfuzz
\raggedbottom

\bibitem{Foner:1989}
S.~Foner, ``Experiments with strong pulsed magnetic fields produced by {Cu/Nb}
  microcomposite wire-wound magnets'',
  \href{https://dx.doi.org/10.1016/0921-4526(89)90455-9}{\emph{Physica B}
  \textbf{155} (1989) 18--22}.

\bibitem{Campbell:1995}
L.~Campbell, Y.~Eyssa, P.~Gilmore, P.~Pernambuco-Wise, D.~G. Parkin, D.~G.
  Rickel, J.~B. Schilligg, and H.~J. Schneider-Muntau, ``The {US 100 T} magnet
  project'',
  \href{https://dx.doi.org/10.1016/0921-4526(94)00943-P}{\emph{Physica B}
  \textbf{211} (1995) 52--55}.

\bibitem{Embury:1998}
J.~D. Embury and K.~Han, ``Conductor materials for high field magnets'',
  \href{https://dx.doi.org/10.1016/S1359-0286(98)80106-X}{\emph{Curr. Opin.
  Solid State Mater. Sci.} \textbf{3} (1998) 304--308}.

\bibitem{Dupouy:1995}
F.~Dupouy, S.~Askenazy, J.~P. Peyrade, and D.~Legat, ``Composite conductors for
  high pulsed magnetic fields'',
  \href{https://dx.doi.org/10.1016/0921-4526(94)00934-N}{\emph{Physica B}
  \textbf{211} (1995) 43--45}.

\bibitem{Pantsyrnyi:2001}
V.~Pantsyrnyi, A.~Shikov, A.~Vorobieva, \emph{et~al.}, ``High strength, high
  conductivity macro- and microcomposite winding wires for pulsed magnets'',
  \href{https://dx.doi.org/10.1016/S0921-4526(00)00741-9}{\emph{Physica B}
  \textbf{294-295} (2001) 669--673}.

\bibitem{Dobatkin:2015}
S.~V. Dobatkin, J.~Gubicza, D.~V. Shangina, N.~R. Bochvar, and N.~Y.
  Tabachkova, ``High strength and good electrical conductivity in {Cu-Cr}
  alloys processed by severe plastic deformation'',
  \href{https://dx.doi.org/10.1016/j.matlet.2015.03.144}{\emph{Mater. Lett.}
  \textbf{153} (2015) 5--9}.

\bibitem{Dong:2020}
L.~Dong, G.~Wei, T.~Cheng, \emph{et~al.}, ``Thermal conductivity, electrical
  resistivity, and microstructure of {Cu/W} multilayered nanofilms'',
  \href{https://dx.doi.org/10.1021/acsami.9b21182}{\emph{ACS Appl. Mater.
  Interfaces} \textbf{12} (2020) 8886--8896}.

\bibitem{Zeng:2016}
L.~F. Zeng, R.~Gao, Q.~F. Fang, X.~P. Wang, Z.~M. Xie, S.~Miao, T.~Hao, and
  T.~Zhang, ``High strength and thermal stability of bulk {Cu/Ta} nanolamellar
  multilayers fabricated by cross accumulative roll bonding'',
  \href{https://dx.doi.org/10.1016/j.actamat.2016.03.034}{\emph{Acta Mater.}
  \textbf{110} (2016) 341--351}.

\bibitem{Sakai:1991}
Y.~Sakai, K.~Inoue, T.~Asano, H.~Wada, and H.~Maeda, ``Development of
  high-strength, high-conductivity {Cu-Ag} alloys for high-field pulsed magnet
  use'', \href{https://dx.doi.org/10.1063/1.105813}{\emph{Appl. Phys. Lett.}
  \textbf{59} (1991) 2965--2967}.

\bibitem{Zhao:2016}
C.~Zhao, Z.~Xiaowei, E.~Wang, R.~Niu, and K.~Han, ``Simultaneously increasing
  strength and electrical conductivity in nanostructured {Cu-Ag} composite'',
  \href{https://dx.doi.org/10.1016/j.msea.2015.11.067}{\emph{Mater. Sci. Eng.
  A} \textbf{652} (2016) 296--304}.

\bibitem{Tsuji:2003}
N.~Tsuji, Y.~Saito, S.-H. Lee, and Y.~Minamino, ``{ARB} (accumulative
  roll-bonding) and other new techniques to produce bulk ultrafine grained
  materials'', \href{https://dx.doi.org/10.1002/adem.200310077}{\emph{Adv. Eng.
  Mater.} \textbf{5} (2003) 338--344}.

\bibitem{Shikov:2001}
A.~Shikov, V.~Pantsyrnyi, A.~Vorobieva, N.~Khlebova, and A.~Silaev, ``High
  strength, high conductivity {Cu-Nb} based conductors with nanoscaled
  microstructure'',
  \href{https://dx.doi.org/10.1016/S0921-4534(01)00109-5}{\emph{Physica C}
  \textbf{354} (2001) 410--414}.

\bibitem{Ding:2021}
C.~Ding, J.~Xu, D.~Shan, B.~Guo, and T.~G. Langdon, ``Sustainable fabrication
  of {Cu/Nb} composites with continuous laminated structure to achieve
  ultrahigh strength and excellent electrical conductivity'',
  \href{https://dx.doi.org/10.1016/j.compositesb.2021.108662}{\emph{Compos. B
  Eng.} \textbf{211} (2021) 108662}.

\bibitem{Carpenter:2014}
J.~S. Carpenter, R.~J. McCabe, S.~J. Zheng, T.~A. Wynn, N.~A. Mara, and I.~J.
  Beyerlein, ``Processing parameter influence on texture and microstructural
  evolution in {Cu-Nb} multilayer composites fabricated via accumulative roll
  bonding'', \href{https://dx.doi.org/10.1007/s11661-013-2162-4}{\emph{Metall.
  Mater. Trans.} \textbf{A45} (2014) 2192--2208}.

\bibitem{Carpenter:2022}
J.~S. Carpenter, C.~Miller, D.~J. Savage, D.~R. Coughlin, E.~L. Tegtmeier, and
  W.~P. Winter, ``The impact of rolling at temperature on conductivity and
  texture in nanolamellar {Cu/Nb} bimetallic composites'',
  \href{https://dx.doi.org/10.1007/s11661-022-06662-w}{\emph{Metall. Mater.
  Trans. A} \textbf{53} (2022) 2208--2213}.

\bibitem{Gu:2017}
T.~Gu, J.-R. Medy, F.~Volpi, \emph{et~al.}, ``Multiscale modeling of the
  anisotropic electrical conductivity of architectured and nanostructured
  {Cu-Nb} composite wires and experimental comparison'',
  \href{https://dx.doi.org/10.1016/j.actamat.2017.08.066}{\emph{Acta Mater.}
  \textbf{141} (2017) 131--141}.

\bibitem{Vidal:2007}
V.~Vidal, L.~Thilly, F.~Lecouturier, and P.-O. Renault, ``Cu nanowhiskers
  embedded in {Nb} nanotubes inside a multiscale {Cu} matrix: {The} way to
  reach extreme mechanical properties in high strength conductors'',
  \href{https://dx.doi.org/10.1016/j.scriptamat.2007.04.001}{\emph{Scr. Mater.}
  \textbf{57} (2007) 245--248}.

\bibitem{Dubois:2012}
J.-B. Dubois, L.~Thilly, P.-O. Renault, and F.~Lecouturier, ``{Cu-Nb}
  nanocomposite wires processed by severe plastic deformation: {Effects} of the
  multi-scale microstructure and internal stresses on elastic-plastic
  properties'', \href{https://dx.doi.org/10.1002/adem.201200033}{\emph{Adv.
  Eng. Mater.} \textbf{14} (2012) 998--1003}.

\bibitem{Rozhnov:2019}
A.~B. Rozhnov, V.~I. Pantsyrny, A.~Kraynev, S.~Rogachev, S.~Nikulin,
  N.~Khlebova, M.~Polikarpova, and M.~Zadorozhnyy, ``Low-cycle bending fatigue
  and electrical conductivity of high-strength {Cu/Nb} nanocomposite wires'',
  \href{https://dx.doi.org/10.1016/j.ijfatigue.2019.105188}{\emph{Int. J.
  Fatigue} \textbf{128} (2019) 105188}.

\bibitem{Ghalehbandi:2019}
S.~M. Ghalehbandi, M.~Malaki, and M.~Gupta, ``Accumulative roll bonding---{A}
  review'', \href{https://dx.doi.org/10.3390/app9173627}{\emph{Appl. Sci.}
  \textbf{9} (2019) 3627}.

\bibitem{Gu2:2019}
T.~Gu, J.-R. Medy, V.~Klosek, \emph{et~al.}, ``Multiscale modeling of the
  elasto-plastic behavior of architectured and nanostructured {Cu-Nb} composite
  wires and comparison with neutron diffraction experiments'',
  \href{https://dx.doi.org/10.1016/j.ijplas.2019.04.011}{\emph{Int. J. Plast.}
  \textbf{122} (2019) 1--30}.

\bibitem{Zhang:2019}
Z.~Zhang, C.~Shao, S.~Wang, X.~Luo, K.~Zheng, and H.~M. Urbassek, ``Interaction
  of dislocations and interfaces in crystalline heterostructures: A review of
  atomistic studies'',
  \href{https://dx.doi.org/10.3390/cryst9110584}{\emph{Crystals} \textbf{9}
  (2019) 584}.

\bibitem{Kacher:2014}
J.~Kacher, B.~P. Eftink, B.~Cui, and I.~M. Robertson, ``Dislocation
  interactions with grain boundaries'',
  \href{https://dx.doi.org/10.1016/j.cossms.2014.05.004}{\emph{Curr. Opin.
  Solid State Mater. Sci.} \textbf{18} (2014) 227--243}.

\bibitem{JWang2:2011}
J.~Wang and A.~Misra, ``An overview of interface-dominated deformation
  mechanisms in metallic multilayers'',
  \href{https://dx.doi.org/10.1016/j.cossms.2010.09.002}{\emph{Curr. Opin.
  Solid State Mater. Sci.} \textbf{15} (2011) 20--28}.

\bibitem{Reiss:1986}
G.~Reiss, J.~Vancea, and H.~Hoffmann, ``Grain-boundary resistance in
  polycrystalline metals'',
  \href{https://dx.doi.org/10.1103/PhysRevLett.56.2100}{\emph{Phys. Rev. Lett.}
  \textbf{56} (1986) 2100--2103}.

\bibitem{Thilly:2000}
L.~Thilly, F.~Lecouturier, G.~Coffe, J.~P. Peyrade, and S.~Askenazy, ``Ultra
  high strength nanocomposite conductors for pulsed magnet windings'',
  \href{https://dx.doi.org/10.1109/77.828466}{\emph{IEEE Trans. Appl.
  Supercond.} \textbf{10} (2000) 1269--1272}.

\bibitem{Spencer:2004}
K.~Spencer, F.~Lecouturier, L.~Thilly, and J.~D. Embury, ``Established and
  emerging materials for use as high-field magnet conductors'',
  \href{https://dx.doi.org/10.1002/adem.200400014}{\emph{Adv. Eng. Mater.}
  \textbf{6} (2004) 290--297}.

\bibitem{Battesti:2018}
R.~Battesti, J.~Beard, S.~B{\"o}ser, \emph{et~al.}, ``High magnetic fields for
  fundamental physics'',
  \href{https://dx.doi.org/10.1016/j.physrep.2018.07.005}{\emph{Phys. Rep.}
  \textbf{765-766} (2018) 1--39},
  \href{https://arxiv.org/abs/1803.07547}{\texttt{arXiv:1803.07547
  [physics.ins-det)]}}.

\bibitem{Beyerlein:2013}
I.~J. Beyerlein, N.~A. Mara, J.~S. Carpenter, \emph{et~al.}, ``Interface-driven
  microstructure development and ultra high strength of bulk nanostructured
  {Cu-Nb} multilayers fabricated by severe plastic deformation'',
  \href{https://dx.doi.org/10.1557/jmr.2013.21}{\emph{J. Mater. Res.}
  \textbf{28} (2013) 1799--1812}.

\bibitem{Nizolek:2016}
T.~Nizolek, I.~J. Beyerlein, N.~A. Mara, J.~T. Avallone, and T.~M. Pollock,
  ``Tensile behavior and flow stress anisotropy of accumulative roll bonded
  {Cu-Nb} nanolaminates'',
  \href{https://dx.doi.org/10.1063/1.4941043}{\emph{Appl. Phys. Lett.}
  \textbf{108} (2016) 051903}.

\bibitem{Hansen:2013}
B.~L. Hansen, I.~J. Beyerlein, C.~A. Bronkhorst, E.~K. Cerreta, and
  D.~Dennis-Koller, ``A dislocation-based multi-rate single crystal plasticity
  model'', \href{https://dx.doi.org/10.1016/j.ijplas.2012.12.006}{\emph{Int. J.
  Plast.} \textbf{44} (2013) 129--146}.

\bibitem{Jia:2013}
N.~Jia, F.~Roters, P.~Eisenlohr, D.~Raabe, and X.~Zhao, ``Simulation of shear
  banding in heterophase co-deformation: {Example} of plane strain compressed
  {Cu-Ag and Cu-Nb} metal matrix composites'',
  \href{https://dx.doi.org/10.1016/j.actamat.2013.04.029}{\emph{Acta Mater.}
  \textbf{61} (2013) 4591--4606}.

\bibitem{Avallone:2019}
J.~T. Avallone, T.~J. Nizolek, T.~M. Pollock, and M.~R. Begley, ``A model for
  high temperature deformation of nanolaminate {Cu-Nb} composites'',
  \href{https://dx.doi.org/10.1016/j.msea.2019.06.026}{\emph{Mater. Sci. Eng.
  A} \textbf{761} (2019) 138016}.

\bibitem{TChen:2020}
T.~Chen, R.~Yuan, I.~J. Beyerlein, and C.~Zhou, ``Predicting the size scaling
  in strength of nanolayered materials by a discrete slip crystal plasticity
  model'', \href{https://dx.doi.org/10.1016/j.ijplas.2019.08.016}{\emph{Int. J.
  Plast.} \textbf{124} (2020) 247--260}.

\bibitem{Shishvan:2021}
S.~S. Shishvan, ``High-temperature tensile and creep behavior of {Cu-Nb}
  composites: {A} discrete dislocation plasticity investigation'',
  \href{https://dx.doi.org/10.1016/j.ijplas.2020.102876}{\emph{Int. J. Plast.}
  \textbf{136} (2021) 102876}.

\bibitem{Tian:2014}
L.~Tian, I.~Anderson, T.~Riedemann, and A.~Russell, ``Modeling the electrical
  resistivity of deformation processed metal-metal composites'',
  \href{https://dx.doi.org/10.1016/j.actamat.2014.06.013}{\emph{Acta Mater.}
  \textbf{77} (2014) 151--161}.

\bibitem{HerveLuanco:2016}
E.~Herv{\'e}-Luanco and S.~Joann{\`e}s, ``Multiscale modelling of transport
  phenomena for materials with {$n$}-layered embedded fibres --- {Part I:
  Analytical} and numerical-based approaches'',
  \href{https://dx.doi.org/10.1016/j.ijsolstr.2016.05.015}{\emph{Int. J, Solids
  Struct.} \textbf{97-98} (2016) 625--636}.

\bibitem{Raabe:1995}
D.~Raabe, ``Simulations of the resistivity of heavily cold worked {Cu-20 wt.\%
  Nb} wires'',
  \href{https://dx.doi.org/10.1016/0927-0256(94)00079-R}{\emph{Comput. Mater.
  Sci.} \textbf{3} (1995) 402--412}.

\bibitem{Lux:1993}
F.~Lux, ``Models proposed to explain the electrical conductivity of mixtures
  made of conductive and insulating materials'',
  \href{https://dx.doi.org/10.1007/BF00357799}{\emph{J. Mater. Sci.}
  \textbf{28} (1993) 285--301}.

\bibitem{Heringhaus:2003}
F.~Heringhaus, H.-J. Schneider-Muntau, and G.~Gottstein, ``Analytical modeling
  of the electrical conductivity of metal matrix composites: application to
  {Ag-Cu and Cu-Nb}'',
  \href{https://dx.doi.org/10.1016/S0921-5093(02)00590-7}{\emph{Mater. Sci.
  Eng. A} \textbf{347} (2003) 9--20}.

\bibitem{Matthiessen:1864}
A.~Matthiessen and C.~Vogt, ``{IV. On} the influence of temperature on the
  electric conducting-power of alloys'',
  \href{https://dx.doi.org/10.1098/rstl.1864.0004}{\emph{Philos. Trans. R. Soc.
  Lond.} \textbf{154} (1864) 167--200}.

\bibitem{Sondheimer:2001}
E.~H. Sondheimer, ``The mean free path of electrons in metals'',
  \href{https://dx.doi.org/10.1080/00018730110102187}{\emph{Adv. Phys.}
  \textbf{50} (2001) 499--537}.

\bibitem{Jin:2013}
Y.~M. Jin, ``Phase field modeling of current density distribution and effective
  electrical conductivity in complex microstructures'',
  \href{https://dx.doi.org/10.1063/1.4813392}{\emph{Appl. Phys. Lett.}
  \textbf{103} (2013) 021906}.

\bibitem{Lu:2004}
L.~Lu, Y.~Shen, X.~Chen, L.~Qian, and K.~Lu, ``Ultrahigh strength and high
  electrical conductivity in copper'',
  \href{https://dx.doi.org/10.1126/science.1092905}{\emph{Science} \textbf{304}
  (2004) 422--426}.

\bibitem{Murashkin:2016}
M.~{\relax Yu}. Murashkin, I.~Sabirov, X.~Sauvage, and R.~Z. Valiev,
  ``Nanostructured al and cu alloys with superior strength and electrical
  conductivity'', \href{https://dx.doi.org/10.1007/s10853-015-9354-9}{\emph{J.
  Mater. Sci.} \textbf{51} (2016) 33--49}.

\bibitem{Chen:2019}
X.~Chen, H.~Zhou, T.~Zhang, \emph{et~al.}, ``{Mechanism of interaction between
  the Cu/Cr interface and its chemical mixing on tensile strength and
  electrical conductivity of a Cu-Cr-Zr alloy}'',
  \href{https://dx.doi.org/10.1016/j.matdes.2019.107976}{\emph{Mater. Des.}
  \textbf{180} (2019) 107976}.

\bibitem{Dingle:1950}
R.~B. Dingle and W.~L. Bragg, ``The electrical conductivity of thin wires'',
  \href{https://dx.doi.org/10.1098/rspa.1950.0077}{\emph{Proc. Royal Soc.
  Lond.} \textbf{A201} (1950) 545--560}.

\bibitem{Fuchs:1938}
K.~Fuchs, ``The conductivity of thin metallic films according to the electron
  theory of metals'',
  \href{https://dx.doi.org/10.1017/S0305004100019952}{\emph{Math. Proc. Camb.
  Philos. Soc.} \textbf{34} (1938) 100--108}.

\bibitem{Gall:2016}
D.~Gall, ``Electron mean free path in elemental metals'',
  \href{https://dx.doi.org/10.1063/1.4942216}{\emph{J. Appl. Phys.}
  \textbf{119} (2016) 085101}.

\bibitem{Mayadas:1970}
A.~F. Mayadas and M.~Shatzkes, ``Electrical-resistivity model for
  polycrystalline films: the case of arbitrary reflection at external
  surfaces'', \href{https://dx.doi.org/10.1103/PhysRevB.1.1382}{\emph{Phys.
  Rev.} \textbf{B1} (1970) 1382--1389}.

\bibitem{Feldman:2010}
B.~Feldman, S.~Park, M.~Haverty, S.~Shankar, and S.~T. Dunham, ``Simulation of
  grain boundary effects on electronic transport in metals, and detailed causes
  of scattering'', \href{https://dx.doi.org/10.1002/pssb.201046133}{\emph{phys.
  stat. sol. (b)} \textbf{247} (2010) 1791--1796},
  \href{https://arxiv.org/abs/0908.2252}{\texttt{arXiv:0908.2252
  [cond-mat.mes-hall]}}.

\bibitem{Karasek:1981}
K.~R. Karasek and J.~Bevk, ``Normal-state resistivity of in situ-formed
  ultrafine filamentary {Cu-Nb} composites'',
  \href{https://dx.doi.org/10.1063/1.329767}{\emph{J. Appl. Phys.} \textbf{52}
  (1981) 1370--1375}.

\bibitem{Karasek:1980}
K.~R. Karasek and J.~Bevk, ``Dislocation resistivity in in situ formed {Cu-Nb}
  multifilamentary composites'',
  \href{https://dx.doi.org/10.1016/0036-9748(80)90340-3}{\emph{Scr. Metall.}
  \textbf{14} (1980) 431--435}.

\bibitem{Brown:1977}
R.~A. Brown, ``Electrical resistivity of dislocations in metals'',
  \href{https://dx.doi.org/10.1088/0305-4608/7/7/026}{\emph{J. Phys. F: Met.
  Phys.} \textbf{7} (1977) 1283--1295}.

\bibitem{Koslowski:2002}
M.~Koslowski, A.~M. Cuiti{\~n}o, and M.~Ortiz, ``A phase-field theory of
  dislocation dynamics, strain hardening and hysteresis in ductile single
  crystals'', \href{https://dx.doi.org/10.1016/S0022-5096(02)00037-6}{\emph{J.
  Mech. Phys. Solids} \textbf{50} (2002) 2597--2635},
  \href{https://arxiv.org/abs/cond-mat/0109447}{\texttt{arXiv:cond-mat/0109447}}.

\bibitem{Beyerlein:2016}
I.~J. Beyerlein and A.~Hunter, ``Understanding dislocation mechanics at the
  mesoscale using phase field dislocation dynamics'',
  \href{https://dx.doi.org/10.1098/rsta.2015.0166}{\emph{Phil. Trans. R. S.}
  \textbf{A374} (2016) 20150166}.

\bibitem{Albrecht:2020}
C.~Albrecht, A.~Hunter, A.~Kumar, and I.~J. Beyerlein, ``A phase field model
  for dislocations in hexagonal close packed crystals'',
  \href{https://dx.doi.org/10.1016/j.jmps.2019.103823}{\emph{J. Mech. Phys.
  Solids} \textbf{137} (2020) 103823}.

\bibitem{Peng:2020}
X.~Peng, N.~Mathew, I.~J. Beyerlein, K.~Dayal, and A.~Hunter, ``A {3D} phase
  field dislocation dynamics model for body-centered cubic crystals'',
  \href{https://dx.doi.org/10.1016/j.commatsci.2019.109217}{\emph{Comput.
  Mater. Sci.} \textbf{171} (2020) 109217},
  \href{https://arxiv.org/abs/1909.10138}{\texttt{arXiv:1909.10138
  [cond-mat.mtrl-sci]}}.

\bibitem{Smith:2020}
L.~T.~W. Smith, Y.~Su, S.~Xu, A.~Hunter, and I.~J. Beyerlein, ``The effect of
  local chemical ordering on {Frank-Read} source activation in a refractory
  multi-principal element alloy'',
  \href{https://dx.doi.org/10.1016/j.ijplas.2020.102850}{\emph{Int. J. Plast.}
  \textbf{134} (2020) 102850}.

\bibitem{Miccoli:2015}
I.~Miccoli, F.~Edler, H.~Pfn{\"u}r, and C.~Tegenkamp, ``The 100th anniversary
  of the four-point probe technique: the role of probe geometries in isotropic
  and anisotropic systems'',
  \href{https://dx.doi.org/10.1088/0953-8984/27/22/223201}{\emph{J. Phys.:
  Cond. Mat.} \textbf{27} (2015) 223201}.

\bibitem{Wenk:2003}
H.-R. Wenk, L.~Lutterotti, and S.~Vogel, ``Texture analysis with the new {HIPPO
  TOF} diffractometer'',
  \href{https://dx.doi.org/10.1016/j.nima.2003.05.001}{\emph{Nucl. Instrum.
  Methods Phys. Res. Sec. A} \textbf{515} (2003) 575--588}.

\bibitem{Vogel:2004}
S.~C. Vogel, C.~Hartig, L.~Lutterotti, R.~B. Von~Dreele, H.-R. Wenk, and D.~J.
  Williams, ``Texture measurements using the new neutron diffractometer {HIPPO}
  and their analysis using the {Rietveld} method'',
  \href{https://dx.doi.org/10.1154/1.1649961}{\emph{Powder Diffr.} \textbf{19}
  (2004) 65--68}.

\bibitem{Carpenter:2012}
J.~S. Carpenter, S.~C. Vogel, J.~E. LeDonne, D.~L. Hammon, I.~J. Beyerlein, and
  N.~A. Mara, ``Bulk texture evolution of {Cu-Nb} nanolamellar composites
  during accumulative roll bonding'',
  \href{https://dx.doi.org/10.1016/j.actamat.2011.11.045}{\emph{Acta Mater.}
  \textbf{60} (2012) 1576--1586}.

\bibitem{Takajo:2018}
S.~Takajo, D.~W. Brown, B.~Clausen, G.~T. Gray, C.~M. Knapp, D.~T. Martinez,
  C.~P. Trujillo, and S.~C. Vogel, ``Spatially resolved texture and
  microstructure evolution of additively manufactured and gas gun deformed
  {304L} stainless steel investigated by neutron diffraction and electron
  backscatter diffraction'',
  \href{https://dx.doi.org/10.1017/S0885715618000350}{\emph{Powder Diffr.}
  \textbf{33} (2018) 141–146}.

\bibitem{Carpenter:2012b}
J.~S. Carpenter, X.~Liu, A.~Darbal, \emph{et~al.}, ``A comparison of texture
  results obtained using precession electron diffraction and neutron
  diffraction methods at diminishing length scales in ordered bimetallic
  nanolamellar composites'',
  \href{https://dx.doi.org/10.1016/j.scriptamat.2012.05.018}{\emph{Scr. Mater.}
  \textbf{67} (2012) 336--339}.

\bibitem{Carpenter:2013}
J.~S. Carpenter, R.~J. McCabe, I.~J. Beyerlein, T.~A. Wynn, and N.~A. Mara, ``A
  wedge-mounting technique for nanoscale electron backscatter diffraction'',
  \href{https://dx.doi.org/10.1063/1.4794388}{\emph{J. Appl. Phys.}
  \textbf{113} (2013) 094304}.

\bibitem{Zeng:2017}
L.~F. Zeng, R.~Gao, Z.~M. Xie, S.~Miao, Q.~F. Fang, X.~P. Wang, T.~Zhang, and
  C.~S. Liu, ``Development of interface-dominant bulk {Cu/V} nanolamellar
  composites by cross accumulative roll bonding'',
  \href{https://dx.doi.org/10.1038/srep40742}{\emph{Sci. Rep.} \textbf{7}
  (2017) 40742}.

\bibitem{You:2021}
C.~You, W.~Xie, S.~Miao, T.~Liang, L.~Zeng, X.~Zhang, and H.~Wang, ``High
  strength, high electrical conductivity and thermally stable bulk {Cu/Ag}
  nanolayered composites prepared by cross accumulative roll bonding'',
  \href{https://dx.doi.org/10.1016/j.matdes.2021.109455}{\emph{Mater. Des.}
  \textbf{200} (2021) 109455}.

\bibitem{Liu:2014}
X.~Liu, N.~T. Nuhfer, A.~D. Rollett, \emph{et~al.}, ``Interfacial orientation
  and misorientation relationships in nanolamellar {Cu/Nb} composites using
  transmission-electron-microscope-based orientation and phase mapping'',
  \href{https://dx.doi.org/10.1016/j.actamat.2013.10.046}{\emph{Acta Mater.}
  \textbf{64} (2014) 333--344}.

\bibitem{CRChandbook}
J.~R. Rumble, ed., \href{https://hbcp.chemnetbase.com}{\emph{CRC Handbook of
  Chemistry and Physics}}, 102nd~ed., (CRC Press, 2021).

\bibitem{Han:1999}
K.~Han, J.~Embury, J.~Sims, L.~Campbell, H.-J. Schneider-Muntau, V.~Pantsyrnyi,
  and A., ``The fabrication, properties and microstructure of {Cu-Ag and Cu-Nb}
  composite conductors'',
  \href{https://dx.doi.org/10.1016/S0921-5093(99)00025-8}{\emph{Mater. Sci.
  Eng. A} \textbf{267} (1999) 99--114}.

\bibitem{Bevk:1978}
J.~Bevk, J.~P. Harbison, and J.~L. Bell, ``Anomalous increase in strength of in
  situ formed {Cu-Nb} multifilamentary composites'',
  \href{https://dx.doi.org/10.1063/1.324573}{\emph{J. Appl. Phys.} \textbf{49}
  (1978) 6031--6038}.

\bibitem{Fiks:1981}
V.~B. Fiks, ``Interaction of conduction electrons with single dislocations in
  metals'', \href{http://jetp.ras.ru/cgi-bin/dn/e_053_06_1209.pdf}{\emph{Sov.
  Phys. JETP} \textbf{53} (1981) 1209}, [\textit{Zh. Eksp. Teor. Fiz.}
  \textbf{80} (1981) 2313--2316].

\bibitem{Antolovich:2004}
S.~D. Antolovich and H.~Conrad, ``The effects of electric currents and fields
  on deformation in metals, ceramics, and ionic materials: An interpretive
  survey'', \href{https://dx.doi.org/10.1081/AMP-200028070}{\emph{Mater. Manuf.
  Process.} \textbf{19} (2004) 587--610}.

\bibitem{Hariharan:2017}
H.~Krishnaswamy, M.~J. Kim, S.-T. Hong, D.~Kim, J.-H. Song, M.-G. Lee, and
  H.~N. Han, ``Electroplastic behaviour in an aluminium alloy and dislocation
  density based modelling'',
  \href{https://dx.doi.org/10.1016/j.matdes.2017.03.072}{\emph{Mater. Des.}
  \textbf{124} (2017) 131--142}.

\end{thebibliography}\endgroup
\end{document}